\documentstyle[aasms4,12pt]{article}

\begin{document}

\title{On the Influence of the Environment in the Star Formation Rates
of a Sample of Galaxies in Nearby Compact Groups}

\vspace{.2in}

\author{J. Iglesias-P\'{a}ramo}
\author{jiglesia@ll.iac.es}
\and
\author{J.M. V\'{\i}lchez}
\author{jvm@ll.iac.es}
\vspace{.2in}
\affil{Instituto de Astrof\'{\i}sica de Canarias\\
C/ V\'{\i}a L\'{a}ctea s.n., 38200 La Laguna, Tenerife, SPAIN}

\vspace{.2in}

\begin{abstract}

We present the results of the study of the Star Formation Rates
(SFRs) of a sample of disk galaxies in nearby compact groups compared
with the SFRs of a sample of field galaxies. For this purpose, H$\alpha$
luminosities and equivalent widths were derived for the galaxies of our
sample. A direct comparison of the equivalent widths and H$\alpha$
luminosities normalized to the $B$ luminosities and estimated area
of the galaxies of both samples yielded the result that the median
values of these quantities were almost identical for both samples,
although the distributions for the compact group sample were broadened
around the mean value with respect to the field galaxy sample. This
result can be explained assuming that though interactions between galaxies
in compact groups can alter their SFRs, the median value of the
normalized SFRs is preserved, being almost indistinguishable from the
corresponding value for field galaxies.

Measuring the global $L_{\rm{H}\alpha}/L_{B}$ of
the groups -- including early-type galaxies -- we find that most of the
groups that show the highest level of $L_{\rm{H}\alpha}/L_{B}$, with respect to
a set of synthetic groups built out of field galaxies, show tidal
features in at least one of their members.
Finally, we have explored the relationship between the
ratio $L_{\rm{H}\alpha}/L_{B}$ and several relevant dynamical
parameters of the groups: velocity dispersion, crossing time, radius
and mass to luminosity ratio, finding no clear correlation. This fact suggests that
the exact dynamical state of a group does not appear to control
the SFR of the group as a whole.

Our results are compatible with a scenario for compact groups of galaxies in which
the dark matter of the group would be arranged in a common halo,
therefore preventing a fast collapse of the galaxies.

\end{abstract}


\section{Introduction}

There has been considerable discussion in the literature about the possibility
that compact groups are chance alignments within looser groups or
clusters, rather than genuinely dense aggregates of galaxies  (Mamon 1995,
Ostriker et al. 1995). The typical median value of the velocity
dispersions of the well known Hickson Compact Groups is about 200
km\,sec$^{-1}$ (Hickson 1982), so that given that this value is of the
order of the typical rotational velocity of the disks of the spirals,
strong effects of the interactions would be expected between their
members. In fact, Mendes de Oliveira \& Hickson (1994) reported
morphological signs of interactions for a large fraction of galaxies
within compact groups, what was interpreted in favor of the bound
system hypothesis. However, Walke \& Mamon (1989) interpreted this
result suggesting that a large number of binaries combined with
projected background galaxies would give the same ratio of interactions
as the one found for the compact groups.

Observations devoted to disentangle the real nature of compact groups
have been carried out by several authors at different wavelength ranges:
Optical (Moles et al. 1994, Pildis et al. 1995), far infrared (Hickson
et al. 1989, Sulentic \& de Mello Raba\c{c}a 1993, Venugopal 1995, Allam
et al. 1996, Verdes-Montenegro et al. 1998), X rays (Ebeling et
al. 1994, Saracco \& Ciliegi 1995, Ponman et al. 1996), radio (Williams
\& Rood 1987, Menon 1995, Huchtmeier 1997). Nevertheless, the problem
still remains unclear.

The study of the Star Formation Rates (SFRs) of the galaxies in compact
groups could also supply information concerning their real nature. It is
known that interactions between disk galaxies can modify their SFRs
under some circumstances of spin alignment, velocity differences and
impact parameter (Mihos et al. 1991). In particular, if the interaction
is so strong to morphologically disturb the disks, an enhancement in the
SFR is expected. This theoretical result has been observationally
confirmed for nearby pairs (Kennicutt et al. 1987, Laurikainen \& Moles
1989) and for samples of peculiar galaxies (e.g. Larson \& Tinsley 1978,
Mazzarella et al. 1991).
Moles et al. (1994), using
broad band photometric data, performed a statistical study on the SFRs of a large
sample of galaxies in compact groups obtaining a slight enhancement
compared to the normal galaxies but with less star formation activity than
paired galaxies. This result suggested that the star formation
properties of compact groups of galaxies
are not dominated by the effects of strong interactions.

However, broad band optical colors are only sensitive to
the SFRs on time scales of the order of
10$^{8}$yr. Better indicators of the evolution of the SFR on shorter
timescales are the FIR luminosities (Telesco 1988, Mazzarella et
al. 1991) and the luminosity of the hydrogen recombination lines
(e.g. Kennicutt 1983). In this work, the SFRs of a sample of galaxies in
compact groups from the ``Catalog of Compact Groups of Galaxies''
(Hickson 1982) were compared with those of a sample of field galaxies in
order to study the influence of the compact group environment on the
SFRs of galaxies. The star formation properties of the most interesting groups have
already been studied in detail in previous papers (Iglesias-P\'{a}ramo \&
V\'{\i}lchez 1997a, 1997b, 1998). Also the complete sample of groups has
been presented in a previous article (V\'{\i}lchez \&
Iglesias-P\'{a}ramo 1998a). In \S 2 we describe our sample of
groups and all the information relevant to the acquisition and reduction
of the data. The observational results of our work together with the
implications on the nature of the compact groups are described in \S 3
and \S 4. Finally, \S 5 contains the conclusions of this work.

\section{Data Reduction and Photometric Results}

The sample of galaxies was previously described by V\'{\i}lchez \&
Iglesias-P\'{a}ramo (1998a), together with their H$\alpha$ images, and
the details of to the acquisition and reduction of the data.

Table~\ref{narrowcal} shows the H$\alpha$ photometry of all the
accordant\footnote{Those galaxies with radial velocity within 1000
$km\, sec^{-1}$ of the median velocity of the group}
galaxies of our sample. The data are corrected for Galactic extinction
following Burstein \& Heiles (1984) and using a standard extinction law (Rieke
\& Lebofsky 1985).
Column (2) shows the Galactic extinction in the $B$ band. Column (3)
shows the absolute $B$ magnitude of the galaxies from de Vaucouleurs et
al.\ (1991). Column (4)
shows the H$\alpha$ luminosity expressed in erg\, sec$^{-1}$. Upper
limits correspond to $3\sigma$ over the sky level. Given that the FWHM
of the filters are about 50\AA\ wide, we could not avoid the
contamination due to the [N{\sc ii}] lines, so hereafter we will refer to
H$\alpha +$[N{\sc ii}] as H$\alpha$. Column (5) shows the
observational uncertainty of the logarithm of the H$\alpha$ luminosity,
computed as the quadratic sum of the observational plus the Poissonian
errors. However, as commented above, there are additional sources of
uncertainty in the H$\alpha$ luminosity: The internal extinction, the
presence of the [N{\sc ii}] lines, the emission of the galactic nuclei and the uncertainty
in the continuum level. Young et al.\ (1996) estimated the total
uncertainty due to these factors at $\approx$20\% of the real value. Column (6) shows
the equivalent width of H$\alpha$ expressed in \AA. The equivalent width
of H$\alpha$ was computed following this formula
\begin{equation}
EW({\rm H}\alpha) = \frac{C_{\alpha}}{C_{cont}} \times W_{f},
\end{equation}
where $C_{\alpha}$ is the number of counts from the galaxy in the net
H$\alpha$ frame, $C_{cont}$ is the number of counts of the galaxy in the
scaled continuum frame and $W_{f}$ is the FWHM of the filter
in \AA. This gives a reasonable estimation of the total H$\alpha$
equivalent width, since we are including the whole galaxy.

Column (7) shows
the present day SFR expressed in solar masses per year. This parameter
only accounts for the stars heavier than $10M_{\odot}$ and has been
computed following Kennicutt (1983):
\begin{equation}
SFR(\geq 10M_{\odot}) = \frac{L(\rm{H}\alpha)}{7.02 \times 10^{41}
\rm{erg\, sec}^{-1}} M_{\odot}\, yr^{-1}
\end{equation}
assuming an IMF of the form:
\begin{equation}
\psi(m) \propto \left\{
\begin{array}{ll}
m^{-1.4} & (0.1 \leq m \leq 1M_{\odot}) \\
m^{-2.5} & (1 \leq m \leq 100M_{\odot})\\
\end{array}
\right.
\end{equation}
Column (8) expresses the confidence level of the measured values for the
H$\alpha$ luminosities: A value of 0 indicates that the galaxy was well
isolated. A value of 1 indicates that the galaxy may have been slightly
contaminated by light from a nearby star or a nearby companion. A value
of 2 means that the flux of the galaxy was strongly contaminated by a
saturated star or that strong difficulties arose in the determination of
the sky. All the data are corrected for Galactic extinction but not for
internal extinction.

Two of the galaxies, HCG44C and HCG92C, are classified as Seyfert
galaxies (Huchra \& Burg 1992). The H$\alpha$ luminosity listed in
Table~\ref{narrowcal} accounts for the contribution due to the whole
galaxy. However, for the subsequent analysis, we reject the
nuclear point source contribution to the H$\alpha$ luminosity of these
galaxies because the ionizing source responsible for the nuclear
H$\alpha$ photons may well be non-thermal. The total nuclear contributions of
HCG44C and HCG92C amount to 68\% and 78\% of the global H$\alpha$
luminosity respectively. Also, HCG68c was found to show LINER-type
emission (Giuricin et al.\ 1990) in the nucleus. The nuclear emission in
this galaxy amounts to 13\% of the total H$\alpha$ emission. 
Galaxies HCG16a and HCG16b have been reported to show a LINER-type spectrum.
However, it was shown by V\'{\i}lchez \& Iglesias-P\'{a}ramo (1998a),
that HCG16a
exhibits a ring of circumnuclear emission but it does not show a nuclear
H$\alpha$ emission region. Thus, we argue that this galaxy could have
been misclassified as a LINER due to the low resolution of the spectrum
analyzed or to the fact that the H$\alpha$ line could be absorbed by the
underlying population of the bulge.
The nuclear emission region in HCG16b
contributes more than 90\% to the total emission of this galaxy.
For the subsequent analysis of the SFRs we will neglect the non-thermal
contributions to the H$\alpha$ luminosities of the galaxies HCG16b, 44c,
68c and 92c.

\section{Observational Results}

\subsection{H$\alpha$ Equivalent Widths}

One of the tools available to study the SFRs of the galaxies in our
sample is the H$\alpha$ equivalent width. It is well known that
this parameter is related to the star formation history of a galaxy,
i.e. to the ratio of the current to past SFR,
and also to the high mass end of the IMF (e.g. Kennicutt et al. 1994 and
references therein).

We have selected the sample of galaxies by Kennicutt \& Kent (1983, hereafter KK83)
to compare the distribution of the H$\alpha$ equivalent widths with the
one for our
sample. This sample of disk galaxies is the largest published with measured
equivalent widths. For our purposes, galaxies belonging to the Virgo
cluster have been removed from the KK83 sample. Also, we have removed
the galaxies that were observed with an aperture smaller than the
diameter of the galaxy. Figure~\ref{ew_kenn} shows the histograms of the H$\alpha$
equivalent widths for both samples. 
The KK83 distribution is narrower than the compact group
one. This effect could be due to the fact that the morphological composition of
both samples are not homogeneous. However the median values of the
distributions are almost
coincident: $\log EW = 1.41$ and $1.45$ for the KK83 and the compact
group samples respectively.

In Figure~\ref{ew} we
have also plotted the distribution of the H$\alpha$ equivalent widths for the
galaxies in our sample binned by Hubble type. The vertical bars indicate
the ranges covered by the galaxies in the KK83 sample for each
bin. The Y axis shows also the birthrate parameter -- defined
in Kennicutt et al. (1994) as a parameterization of the star formation
history of a disk galaxy --.
There are no significant differences between both samples,
excepting the extreme H$\alpha$ equivalent width measured for the
Im merging galaxy in HCG31.

\subsection{Present Day SFRs}

As has been shown in the previous section, no statistical differences
in the star formation history between the compact groups sample and the
KK83 sample were found from their H$\alpha$ equivalent widths. In this
section we are trying to shed light in the possible influence of the
environment on the present day SFRs of the galaxies in compact groups
as compared to a sample of field galaxies. For this purpose we have
built an extended sample of disk galaxies from several sources in the
literature: Young et al. (1996), KK83, Hunter \& Gallagher (1986) and
Miller \& Hodge (1994). For galaxies with more than one estimation
of the H$\alpha$ flux, the most recent was chosen.

The distances of the galaxies were derived from the Kraan-Korteweg
Catalogue (1986), or assuming a pure Hubble flow with $H_{0} = 100 km\,
sec^{-1}\, Mpc^{-1}$ for those galaxies not included in this catalogue.
The H$\alpha$ luminosities and absolute magnitudes were calculated
using these distances. The data were corrected for Galactic
extinction following Burstein \& Heiles (1984). No further correction
was applied to the H$\alpha$ fluxes so that we could compare them
directly to our own data.  However, as the amount of internal
extinction is not expected to present large variations from a mean
value for the two samples and taking into account that the main
conclusions of our work are statistical, we claim that the effect of internal
extinction will hardly affect the results of this paper.

We have discarded from the field sample those galaxies belonging to
the Virgo Cluster and to any of the Abell clusters as well as paired galaxies,
merged galaxies and galaxies presenting nearby
satellites. Thus, our field sample is free of environmental effects.
The fractional abundance of galaxies of each morphological type is
very similar to the compact groups sample, so that any
effect that depends on the composition of the samples is excluded.
Table~\ref{compsamp} contains the main observational properties of the
galaxies of the field sample. Columns (1) to (5) contain selected
properties of the galaxies.

Given that the compact group and the field samples cover a large range
in magnitudes and masses, we have normalized the H$\alpha$ luminosities to the $B$
luminosity in order to compare their current SFRs. Figure~\ref{hab3}
shows $L_{\rm{H}\alpha}$ against the absolute $B$ magnitude for the
galaxies of our sample -- open squares -- and the field sample --
asterisks --. As this figure
shows, no galaxies fainter than $M_{B} = -14$ are present in the compact
group sample. This bias is mostly due to the
selection criterion for compact groups which limits the magnitude
difference between two galaxies in the group to 3 magnitudes. However,
the high luminosity limits are similar for the two samples. Also, it can
be seen that for magnitudes brighter than $M_{B} = -17$, the scatter in
the distribution of the compact group galaxies is larger than for the
field galaxies.

Figure~\ref{nhisto1} shows the corresponding
histograms of $L_{\rm{H}\alpha}/L_{B}$ for both samples (the upper
histogram shows the distribution for the compact group sample and the
lower one shows the distribution for the field sample). The irregular galaxies of the two samples have been
highlighted with grey bins. It can be seen that the irregular galaxies
from the compact group sample extend over one order of magnitude
whereas the irregulars from the field sample extend over almost two
orders of magnitude in $L_{\rm{H}\alpha}/L_{B}$.
This effect is reflecting the selection criterion mentioned above for
Hickson compact groups. Thus, no low surface brightness irregulars are present in
the compact groups of our sample.
Both distributions show a
maximum around $\log L_{\rm{H}\alpha}/L_{B} \approx -3$. The
histogram of compact groups also shows a secondary maximum at around $\log
L_{\rm{H}\alpha}/L_{B} \approx -4$ which is due to the existence of
spiral galaxies with a low level of H$\alpha$ emission. However,
although this secondary maximum cannot be claimed with a great
statistical significance because of the small number of data points, it
is clear that this
feature does not have a counterpart in the histogram of field
galaxies. The highest values of $\log L_{\rm{H}\alpha}/L_{B}$ are also
found in the compact group distribution. The median
values of $\log L_{\rm{H}\alpha}/L_{B}$ are $-2.85$ for the sample of
galaxies in compact
groups and $-2.92$ for the field sample. These values are not
significantly different given the overall errors in
$L_{\rm{H}\alpha}/L_{B}$.
The significance level obtained was
0.57 after applying the Kolmogorov-Smirnoff test to both distributions.

A further criterium to compare the SFRs of different samples of galaxies is
the SFR per unit surface area of the galaxies. Given that for disk galaxies most
of the SFR is concentrated in the disk, this indicator should be a better
indicator of the level of star formation, since for the earlier spirals
the blue luminosity has a strong contribution from the bulge, where very
little star formation activity normally occurs.
Figure~\ref{sfrarea} shows the
histogram of the present day SFRs per unit area for the galaxies of the
compact group sample (a) and for the field sample (b). The grey bins in
the plot correspond to the irregulars in both samples. The
areas of the galaxies were parameterized according to the product of the lengths
of the major and minor axes following de Vaucouleurs et al.\ (1991). The
maxima of the distributions are slightly displaced: $\log SFR/Area =
-8.9M_{\odot}\, yr^{-1}\, pc^{-2}$ for the compact group sample and
$\log SFR/Area = -8.3M_{\odot}\, yr^{-1}\, pc^{-2}$ for the field
sample. Moreover, the distribution corresponding to the compact group
galaxies is less symmetric and broader than that of the field galaxies
and seems to be weighted towards low SFRs per unit area. The
distribution of the irregular galaxies in the two samples are quite
different: It covers a range of about 3 orders of magnitude for the
field galaxies whereas it is restricted to the high SFRs per unit
surface area for the compact group galaxies, again due to the selection
criterion mentioned above. The median values of the
distributions are $\log SFR/Area = -8.66M_{\odot}\, yr^{-1}\, pc^{-2}$
for the compact group sample and $\log SFR/Area = -8.42M_{\odot}\,
yr^{-1}\, pc^{-2}$ for the field sample. The significance level obtained was
0.50 after applying the Kolmogorov-Smirnoff test.
The present day SFR per unit surface area was found to be enhanced in
samples of interacting
galaxies (Bushouse 1987, Kennicutt et al.\ 1987, Laurikainen \& Moles
1989). However, the same result does not hold for
the galaxies in our compact group sample.
Concerning the shape of the distributions, the
sample of compact groups shows an excess of low H$\alpha$ luminosity spirals which
does not have a counterpart in the distribution for the field galaxies.

The main result found in this section is that there is no global
enhancement in the star formation histories of the galaxies in the
compact groups sample neither in their present day SFRs, when compared
to galaxies of the field sample. This suggests that although
interactions among galaxies in compact groups are expected to occur,
their characteristic signature should be different from a straight
enhancement of the SFR of the galaxies involved.

\subsection{Total H$\alpha$ Luminosity of the Compact Groups}

In this section we analyze the total H$\alpha$ luminosity
of the groups normalized to their $B$ luminosity. In the previous
section we have found that the individual disk galaxies in our
sample do not show a significant enhancement in the present day SFR
with respect to the galaxies in our field sample.
An enhancement in the normalized H$\alpha$ luminosity was found,
however, for the early-type galaxies of the compact group sample (see
V\'{\i}lchez \& Iglesias-P\'{a}ramo 1998b) compared to the sample of
field early-type galaxies. This enhancement was attributed to the
accretion of gas by the early-type galaxies in the groups from the outer
envelopes of gas-rich galaxies during close passages.
In this section we will test whether there is any
environmental effect related to the dynamical state of the groups which
affects the total H$\alpha$ emission of a given group.

For this purpose, we built synthetic groups composed of galaxies taken from the
field sample. We have added to our field sample the sample of field ellipticals and
lenticulars listed in V\'{\i}lchez \& Iglesias-P\'{a}ramo (1998b) in
order to cover all the morphological types.
For each real group of our sample a set of synthetic
groups was constructed with the same number of galaxies and fractional
abundance of morphological types as the real group, and restricted by
the condition that the maximum difference in absolute magnitude of the
members of the synthetic groups is three, following the selection
criterion imposed by Hickson (1982). Using these criteria, for
each real group of our sample we selected all the possible synthetic
groups of galaxies from the field sample.
Then, the ratio $\log
L_{\rm{H}\alpha}/L_{B}$ was computed for all the groups. The final
result is plotted in Figure~\ref{nhisto}. This Figure shows the ratio
$\log L_{\rm{H}\alpha}/L_{B}$ for the groups of our sample represented
by open squares. The median value of this ratio for the corresponding
synthetic groups is indicated by open triangles. Vertical error bars
correspond to one standard deviation in $\log L_{\rm{H}\alpha}/L_{B}$.

As can be seen, 6 of the groups -- HCG16, 31, 54, 92, 93 and 100 -- show
a $L_{\rm{H}\alpha}/L_{B}$ ratio higher than $1\sigma$ above the median
values of the distribution of synthetic groups, 4 of the groups -- HCG7,
30, 44 and 61 -- show a $L_{\rm{H}\alpha}/L_{B}$ ratio lower than
$1\sigma$ below the median values and 6 of the groups -- HCG2, 23, 37,
68, 79 and 95 -- lie within $1\sigma$ of the median values
of the synthetic groups. Of the 6 groups that lie more than $1\sigma$
above the median values of the synthetic groups, 5 of them -- HCG16, 31,
54, 92 and 100 -- lie even above the highest value reached by any of the
corresponding synthetic groups. These 5 groups are composed by nearly 100\% of
spirals or irregulars. On the contrary, all those groups that lie lower
than $1\sigma$ below the median values contain a lower fraction of
late-type galaxies.

Four of the groups of our sample have a first-ranked\footnote{Brightest
galaxy in the group} elliptical --
HCG37, 79, 93 and 95 -- whereas two of them have a first-ranked S0 --
HCG61 and 68 --. These groups do not show any preference for a
value of $L_{\rm{H}\alpha}/L_{B}$ above or below the median value of the
synthetic groups. Thus, although
the ellipticals in the groups show a higher value of
$L_{\rm{H}\alpha}/L_{B}$ than the field ellipticals, the presence of a
bright elliptical in the group is not the key factor controlling
the total $L_{\rm{H}\alpha}/L_{B}$ of the group.

There are 4 groups in the compact group sample that contain interacting
pairs showing clear tidal tails: HCG16, 31, 92 and 95. Three of these
groups show $L_{\rm{H}\alpha}/L_{B}$ $1\sigma$ above the
median value of the synthetic groups. Thus it seems that when
interactions are so strong to disrupt the disks of the galaxies in
such a way that tidal features are developed, then a clear enhancement
in the present day SFR of at least one of the galaxies involved is
observed.

It can be expected that the H$\alpha$ emission of the groups may be
related to the gaseous content of the group. Several authors have
studied the H{\sc i} content of the HCGs and have generated very recently an
almost complete database of the catalogue. The H{\sc i} masses for all the
groups of our sample are 
available in the literature (see Williams \& Rood 1987,
Huchtmeier 1997). With these published data we plotted the ratio
$L_{\rm{H}\alpha}/L_{B}$ against $M$(\rm{H}{\sc i})/$L_{B}$ for the groups of
our sample in Figure~\ref{havhi}, and we found a
slight correlation in the expected sense that the H$\alpha$ luminosity
increases with the H{\sc i} content (both relative to the total $B$
luminosity). However, the data show a large scatter for the groups with a
higher content of H{\sc i}. This means that H{\sc i} is necessary
for a high SFR, but some groups with a high level of H{\sc i} show a low
value of $L_{\rm{H}\alpha}/L_{B}$ and thus a low SFR.

The molecular gas content of disk galaxies is known to be a
better indicator of the present day star formation activity of
galaxies. Perea et al. (1997) carried out a program devoted to study the
relationship between the quantity of molecular gas and the $B$
luminosity for isolated galaxies and they found a nonlinear dependence
between these two quantities for spiral galaxies spanning 4 orders of
magnitude. Furthermore, Verdes-Montenegro et al. (1998) found that the
CO and FIR properties of a distance limited complete sample of Hickson
compact group galaxies were surprisingly similar to isolated
spirals. This result agrees well with our finding that the present day
SFRs of the compact group galaxies are on average indistinguishable from
the one corresponding to the field galaxies.

As compact groups are supposed to be dynamical entities, and given that
some interactions are known to enhance the SFRs of the disk galaxies, it
could be thought that some connection between the dynamical properties
of the groups and the SFRs of the member galaxies is expected.
In order to explore any possible relationship between the global SFR of the groups
and their dynamical state, we have plotted
the ratio $L_{\rm{H}\alpha}/L_{B}$ against four
different dynamical parameters of the groups.
Figure~\ref{havel} shows the ratio $L_{\rm{H}\alpha}/L_{B}$ plotted
against (a) the velocity dispersion, (b) the dimensionless crossing time, (c) the median
projected separation and (d) the mass-to-light ratio. No correlation
with the $L_{\rm{H}\alpha}/L_{B}$ luminosities was found in any of
the four plots.

As plot (a) shows, the range in velocity dispersion covered by the
groups in our sample is around 200 km\, sec$^{-1}$, that is of the order
of the typical
values for the rotation velocity of disk galaxies. This fact implies that the
damage to be produced by the interactions should be maximal under these
conditions. However, the lack of correlation with the normalized
H$\alpha$ luminosity means that some other parameter governing the interaction
plays a role in the regulation of the star formation of galaxies.
The range of crossing times covered by the groups of our sample is
very extended, as plot (b) shows, and again no correlation was found
with the normalized H$\alpha$ luminosity. The number of interactions is
related to the crossing time in the sense that the larger the crossing
time, the less interactions occur.
Thus, this result means that the number of interactions that a galaxy is
susceptible to experience is not an important factor that controls the
SFRs of the galaxies in the groups.
The median projected separation of the groups are shown in plot (c). As
this plot shows, there is no correlation between this parameter and the
normalized H$\alpha$ luminosity. This result suggests that the impact
parameter alone does not regulate the star
formation induced by interactions.
Finally, plot (d) shows the mass to luminosity ratio of the groups
against the normalized H$\alpha$ luminosity. The absence of correlation
between these two quantities means that the total amount of dark matter is not
a key parameter controlling the SFRs of the galaxies in the group.
The overall result of this analysis is that the dynamical parameters
considered above have no influence on the SFR of the system as a
whole.

Thus, in this section we have seen that the total normalized H$\alpha$
luminosities of the compact groups are on average very similar to the
computed for the sample of synthetic groups of field galaxies. Only
when interactions are strong so as to develop tidal tails,
the level of H$\alpha$ is enhanced in the compact groups with respect to
the synthetic ones.

\section{Discussion}

In the previous sections, we have shown that the present day SFRs of the
disk galaxies of our sample of compact groups are not enhanced on
average with respect to the values measured for the field sample. But,
the distribution of normalized H$\alpha$ luminosities is broadened
compared to the corresponding to the field sample, thus meaning that the
group environment is modifying the star formation properties of the
galaxies in the groups. In particular, when tidal features are present,
the galaxies in the groups tend to present the highest values of
their present day SFR.

As it was demonstrated by the study by Mendes de Oliveira
and Hickson (1994), and it would be easily inferred from the
values of the relevant dynamical parameters of the groups listed
in the previous section, it is clear that galaxy interactions  
are ubiquitous in most of the HCGs. Thus, from the environmental 
point of view, interactions are expected to be one of the most
efficient agents affecting star formation in group galaxies.
However, what appears not straightforward is the prediction of 
the actual effect of interactions on the present day SFR. Published 
simulations on induced star formation and interactions (Olson \& Kwan 1990,
Mihos et al. 1991) predict that only in some cases their SFR is expected
to be enhanced during this process, whereas the same simulations predict
that the SFR could even be depleted for given combinations of parameters
describing the interaction. The results found in this work agree with
these predictions, in the sense that we can explain the broadening of
the normalized H$\alpha$ luminosity as the result of the modification of
the SFR due to interactions. The galaxies very
luminous in H$\alpha$ present in some of our groups would correspond to
the privileged cases of interaction in which the SFR is enhanced.

Our result also agrees with the previous finding by Hashimoto et al. (1998)
who found that the highest levels of star formation are more prevalent
in the intermediate environment of poor clusters than in either the
field or rich clusters. Tidal features produced during interactions are
difficult to occur in rich clusters because their velocity dispersions
are around 750 km/sec$^{-1}$ (Zabludoff et al. 1990), that is, several
times higher than for compact groups, thus preventing disrupting
interactions between galaxies. For the field galaxies, interactions are
not likely to happen because of the lack of neighbor galaxies.

Further information can be extracted combining the H{\sc i} content of the
spirals and
the H$\alpha$ luminosities of the early-type galaxies in our sample of
compact groups. Williams \& Rood (1987) and Huchtmeier (1997) reported
that most of the spirals in the Hickson compact groups are H{\sc i}
deficient. This result can be explained assuming that most interactions
between galaxies just produce harassment in their external halos. The
depletion of the gaseous halos would result in no changes or an inhibition of the SFR
of the galaxies. The gas lost by the disk galaxies could be accreted by
the more massive ellipticals and lenticulars, where it could result in
some new extra star formation after compression during the accretion
process. This would explain the enhancement in the H$\alpha$ luminosity
found for the early-type galaxies of this sample (V\'{\i}lchez \&
Iglesias-P\'{a}ramo, 1998b). However, in order to set this hypothesis,
more observational data of early-type galaxies are required.

Concerning the dynamical nature of the compact groups of galaxies, the order
of magnitude of their lifetimes still remains unclear: Barnes (1989)
showed that if we assume that the dark matter is attached to the
individual galaxies, the lifetimes of compact groups of galaxies should
be no longer than few crossing times. The final fate of the groups would
be a massive elliptical galaxy originated by merging of the initial
galaxies. But, if this were the case, we should see more merger remnants
with strongly enhanced SFRs instead of a majority of normal groups
showing present day SFRs similar to the field galaxies, as it is actually observed.
This problem can be sorted out assuming that the bulk of dark matter is
mostly arranged in a central halo and shared by all the galaxies in the
group. The recent simulations by G\'{o}mez-Flechoso (1997) indicate that
if the galaxies are immersed in a dark matter envelope, the
lifetime of the group could be as large as $10^{10}$yr without substantial changes
in the structure of the galaxies except in their outer parts. She
made numerical simulations of a compact group of four
galaxies, assuming that dark matter was mostly placed in a common halo
and assuming that the galaxies are aggregates of particles with velocity
dispersions between 100 and 200 km\, sec$^{-1}$. She found that the
sizes of the galaxies are reduced by less than 30\% after
$10^{10}$yr and that the aspect of the groups remained almost similar to
the initial one. The assumption that galaxies are composed systems, rather
than point particles, coupled with a new mathematical formulation of
the dynamical friction can enlarge the timescales as long as double of the
classical estimations. 

Finally, assuming that the lifetimes of compact groups are so large, we still can
address the question about the morphological transformation of spirals
into spheroidals in clusters of galaxies, proposed by Moore et
al. (1998). These authors suggest that after approximately 5 Gyr, the
spirals in clusters transform into spheroidal systems due to encounters
with bright galaxies and the cluster's tidal field. The transformation
is mostly due to the cumulative effect of encounters at speeds of
several thousands of km\, sec$^{-1}$. However, the typical velocity
dispersions in compact groups, of the order of 200 km\, sec$^{-1}$,
make the probability of a high velocity encounters very low, and thus
there is little room for galaxy transformation via harassment. In fact,
although it is known that the fraction of spirals in compact groups is
lower than in the field (Hickson 1982), till the date no faint
spheroidals have been reported in compact groups.

\section{Conclusions}

In this paper we have studied the effects of the environment on the SFRs
of a sample of disk galaxies in compact groups, using H$\alpha$
equivalent widths and luminosities to derive their star formation
histories and present day SFRs respectively. A direct comparison
with a sample of field galaxies has yielded the following results:

The present day SFRs of the galaxies in compact groups are slightly modified by the
environment, but there is not an overall enhancement with respect to the
field galaxies. This finding agrees with previous results from
theoretical simulations
that only some kind of interactions are able to enhance the SFR of the
galaxies involved, whereas the rest of them keep it unchanged or even
tend to inhibit it.

The total H$\alpha$ luminosity of the groups relative to their $B$
luminosity is not significantly enhanced compared to the values of a
sample of synthetic groups of field galaxies. In fact, most of the
groups show normalized H$\alpha$ luminosities below or equal to the ones shown by
their corresponding synthetic groups.
The $L_{\rm{H}\alpha}/L_{B}$ ratio is slightly correlated with
the $M$(H{\sc i})/$L_{B}$ ratio in the sense that the groups that show a
high level of $L_{\rm{H}\alpha}/L_{B}$ also have a high content of
H{\sc i}. However, some of the groups show a high content of H{\sc i} but quite a
low level of $L_{\rm{H}\alpha}/L_{B}$. 
No clear correlations were found
between the total $L_{\rm{H}\alpha}/L_{B}$ ratio of the groups and
several relevant dynamical parameters, a fact that suggests that the
exact dynamical state of
the groups does not control the SFRs of the member galaxies.

Finally, our results appear compatible with a scenario for compact groups of
galaxies in which galaxies are embedded in a dark matter halo that enlarges the
lifetime of the group, preventing the galaxies for rapid merging and collapse,
what would result in significantly enhanced SFRs for the observed
compact groups.

\acknowledgments

We want to acknowledge to John Beckman for careful reading of the last
version of this document and for interesting comments and
suggestions. Thanks must also be given to Mari\'{a}ngeles
G\'{o}mez-Flechoso for interesting comments about dynamical friction.
The INT and the JKT are operated on the island of La Palma by the RGO in
the Spanish Observatorio del Roque de Los Muchachos of the Instituto de
Astrof\'{\i}sica de Canarias.
The 2.2m Telescope is operated by the MPIA in
the Spanish Observatorio de Calar Alto.
This research has made use of the NASA/IPAC Extragalactic Database (NED)
which is operated by the Jet Propulsion Laboratory, California Institute
of Technology, under contract with the National Aeronautics and Space
Administration.

\newpage

\figcaption{Histograms of the H$\alpha$ equivalent widths for the Kennicutt \& Kent
(1983) and the compact groups sample.\label{ew_kenn}}

\figcaption{Distribution of H$\alpha$ equivalent widths for the galaxies of
our sample binned by Hubble type. The vertical bars indicate the range
covered by the KK83 sample.\label{ew}}

\figcaption{H$\alpha$ luminosity against absolute $B$ magnitude for the
spiral and irregular galaxies in our sample -- open squares -- and the
spiral and irregular galaxies of the
comparison sample -- small asterisks --.\label{hab3}}

\figcaption{Distributions of $L_{\rm{H}\alpha}/L_{B}$ for the compact group
-- upper plot --  and the field -- lower plot -- samples.\label{nhisto1}}

\figcaption{Distribution of the SFR per unit area for the galaxies in the
sample of compact groups (a) and in the comparison sample
(b). The filled bins correspond to the
irregular galaxies in both samples.\label{sfrarea}}

\figcaption{Comparison of the ratio $L_{\rm{H}\alpha}/L_{B}$ of the groups of our
sample with the average values for the synthetic groups. Open squares indicate the measured
values for our sample whereas open triangles indicate the average values
for the synthetic groups. Vertical error bars correspond to one standard
deviation. \label{nhisto}}

\figcaption{$\log L_{\rm{H}\alpha}/L_{B}$ against $\log M$(H{\sc i})/$L_{B}$
for the groups of our sample.\label{havhi}}

\figcaption{$L_{\rm{H}\alpha}/L_{B}$ of the groups against the velocity
dispersion (a), the dimensionless crossing time (b), the median
projected separation (c) and the mass-to-light ratio (d).\label{havel}}

\newpage

\begin{deluxetable}{lccrccrc}
\tablecaption{Characteristic parameters of the galaxies: (1) Galaxy
identifier. (2) Galactic $B$ band extinction. (3) $B$ magnitude from
de Vaucouleurs et al.\ (1991). (4) $\log$
of the H$\alpha$ luminosity in $erg\, sec^{-1}$. Upper limits are $3\sigma$ over
the sky level. (5) Photometric
uncertainty in the $\log L(\rm{H}\alpha)$. (6) H$\alpha$ equivalent width
in \AA. (7) $\log$ of the SFR($M_{*} \geq
10M_{\odot}$) in $M_{\odot} yr^{-1}$. (8) Confidence level of the
measured H$\alpha$ luminosity: 0 good, 1 regular, 2 bad. All the data
are corrected for Galactic extinction.\label{narrowcal}}
\tablehead{
\colhead{Gal.} & 
\colhead{$A_{B}$} & 
\colhead{$M_{B}$} & 
\colhead{$\log L_{H\alpha}$} & 
\colhead{$\delta \log L_{H\alpha}$} & 
\colhead{$EW_{\rm{H}\alpha}$} & 
\colhead{$\log$ SFR} &
\colhead{Conf.}
}
\startdata
2a & 0.14 & -19.30 & 41.24 & 0.03 & 89 & $-0.61$ & 0 \\
2b & 0.14 & -18.70 & 41.11 & 0.03 & 101 & $-0.74$ & 0 \\
2c & 0.14 & -18.10 & 40.29 & 0.03 & 49 & $-1.56$ & 0 \\
7a & 0.00 & -19.71 & 40.75 & 0.03 & 14 & $-1.10$ & 0 \\
7b & 0.00 & -19.33 & $<36.53$ & --- & --- & $<-5.32$ & 0  \\
7c & 0.00 & -19.56 & 40.60 & 0.03 & 25 & $-1.25$ & 0 \\
7d & 0.00 & -18.33 & 40.06 & 0.03 & 24 & $-1.79$ & 0 \\
16a & 0.00 & -20.08 & 41.19 & 0.03 & 27 & $-0.66$ & 0 \\
16b & 0.00 & -19.30 & 40.42 & 0.03 & 8 & $-1.43$ & 0 \\
16c & 0.00 & -19.42 & 41.50 & 0.03 & 97 & $-0.35$ & 0 \\
16d & 0.00 & -19.06 & 40.82 & 0.03 & 37 & $-1.03$ & 0 \\
23a & 0.26 & -18.43 & 40.21 & 0.03 & 7 & $-1.64$ & 0 \\
23b & 0.26 & -18.42 & 40.66 & 0.03 & 14 & $-1.19$ & 0 \\
23c & 0.26 & -17.59 & 39.99 & 0.03 & 6 & $-1.86$ & 0 \\
23d & 0.26 & -17.12 & 40.66 & 0.03 & 29 & $-1.19$ & 0 \\
30a & 0.14 & -19.67 & 39.69 & 0.03 & 3 & $-2.16$ & 1 \\
30b & 0.14 & -19.13 & 39.69 & 0.03 & 3 & $-2.16$ & 0 \\
30c & 0.14 & -17.59 & 39.02 & 0.04 & 11 & $-2.83$ & 0 \\
30d & 0.14 & -17.13 & $<36.63$ & --- & --- & $<-5.22$ & 0 \\
31a\tablenotemark{a} & 0.24 & -18.44 & 41.66 & 0.03 & 275 & $-0.19$ & 1 \\
31b                  & 0.24 & -17.93 & 40.71 & 0.03 & 97 & $-1.14$ & 0 \\
31c\tablenotemark{a} & 0.24 & -18.44 & 41.66 & 0.03 & 275 & $-0.19$ & 1 \\
31g\tablenotemark{b} & 0.24 & -17.33 & 40.56 & 0.03 & 99 & $-1.29$ & 0 \\
37a & 0.03 & -20.85 & 41.04 & 0.03 &  16 & $-0.81$ & 1 \\
37b & 0.03 & -19.03 & 40.60 & 0.03 &  19 & $-1.25$ & 0 \\
37c & 0.03 & -17.29 & 40.13 & 0.03 &  16 & $-1.72$ & 0 \\
37d & 0.03 & -17.95 & 40.25 & 0.03 &  61 & $-1.60$ & 0 \\
37e & 0.03 & -17.83 & 39.69 & 0.04 &  20 & $-2.16$ & 0 \\
44a & 0.09 & -18.70 & 39.63 & 0.05 & 4 & $-2.22$ & 2 \\
44b & 0.08 & -18.64 & 39.30 & 0.05 & 2 & $-2.55$ & 1 \\
44c & 0.09 & -17.60 & 39.78 & 0.05 & 4 & $-2.07$ & 0 \\
44d & 0.08 & -16.62 & 40.20 & 0.04 & 69 & $-1.65$ & 1 \\
\enddata
\tablenotetext{a}{A single value has been applied for HCG31a and HCG31c.}
\tablenotetext{b}{This galaxy was not originally listed as a member of
the group.}
\end{deluxetable}

\addtocounter{table}{-1}

\begin{deluxetable}{lccrccrc}
\tablecaption{Continued.}
\tablehead{
\colhead{Gal.} & 
\colhead{$A_{B}$} & 
\colhead{$M_{B}$} & 
\colhead{$\log L_{H\alpha}$} & 
\colhead{$\delta \log L_{H\alpha}$} & 
\colhead{$EW_{\rm{H}\alpha}$} & 
\colhead{$\log$ SFR} &
\colhead{Conf.}
}
\startdata
54a & 0.00 & -15.40 & 39.04 & 0.04 & 11 & $-2.81$ & 1 \\
54b & 0.00 & -14.50 & 39.93 & 0.04 & 136 & $-1.92$ & 1 \\
54c & 0.00 & -14.00 & 39.12 & 0.04 & 55 & $-2.73$ & 1 \\
54d & 0.00 & -13.08 & 39.49 & 0.04 & 88 & $-2.36$ & 1 \\
61a & 0.02 & -19.62 & 40.11 & 0.04 & 5 & $-1.74$ & 0 \\
61c & 0.02 & -18.65 & 40.18 & 0.04 & 11 & $-1.67$& 0  \\
61d & 0.02 & -18.48 & 39.37 & 0.04 & 3 & $-2.48$ & 0 \\
68a & 0.00 & -19.78 & 40.82 & 0.04 & 6 & $-1.03$ & 2 \\
68b & 0.00 & -19.04 & 40.48 & 0.04 & 5 & $-1.37$ & 2 \\
68c & 0.00 & -19.21 & 41.01 & 0.04 & 26 & $-0.84$ & 2 \\
68d & 0.00 & -17.48 & 39.50 & 0.04 & 4 & $-2.35$ & 1 \\
68e & 0.00 & -17.06 & 39.47 & 0.04 & 5 & $-2.38$ & 1 \\
79a & 0.11 & -18.24 & 39.77 & 0.05 & 6 & $-2.08$ & 1 \\
79b & 0.11 & -18.73 & 39.79 & 0.05 & 4 & $-2.06$ & 1 \\
79c & 0.11 & -17.65 & $<36.45$ & --- & --- & $<-5.40$ & 1 \\
79d & 0.11 & -16.94 & 39.30 & 0.05 & 13 & $-2.55$ & 1 \\
92b\tablenotemark{a} & 0.33 & -19.81 & 41.81 & 0.03 & 36 & $-0.04$ & 1 \\
92c                  & 0.33 & -19.89 & 41.66 & 0.03 & 30 & $-0.19$ & 0 \\
92d\tablenotemark{a} & 0.33 & -19.73 & 41.81 & 0.03 & 36 & $-0.04$ & 1 \\
92e                  & 0.33 & -18.95 & $<37.50$ & --- & --- & $<-4.35$ & 0 \\
93a & 0.18 & -20.01 & 40.53 & 0.03 & 3 & $-1.32$ & 0 \\
93b & 0.18 & -19.44 & 41.40 & 0.03 & 28 & $-0.45$ & 0 \\
93c & 0.18 & -18.82 & 40.27 & 0.03 & 3 & $-1.58$ & 0 \\
93d & 0.18 & -17.76 & $<37.17$ & --- & --- & $<-4.68$ & 0 \\
95a & 0.13 & -20.37 & 40.51 & 0.05 & 6 & $-1.34$ & 0 \\
95c & 0.13 & -19.57 & 41.10 & 0.05 & 31 & $-0.75$ & 0 \\
95d & 0.13 & -18.51 & 40.22 & 0.05 & 6 & $-1.63$ & 0 \\
100a                  & 0.14 & -19.50 & 41.43 & 0.03 & 27 & $-0.42$ & 0 \\
100b                  & 0.14 & -18.32 & 40.97 & 0.03 & 48 & $-0.88$ & 0 \\
100c                  & 0.14 & -17.68 & 40.50 & 0.03 & 32 & $-1.35$ & 0 \\
100d\tablenotemark{b} & 0.14 & -16.81 & 40.28 & 0.03 & 34 & $-1.57$ & 0 \\
\enddata
\tablenotetext{a}{A single value has been applied for the H$\alpha$
luminosity of HCG92b and HCG92d.}
\tablenotetext{b}{A value of 5300 km\, sec$^{-1}$ has been assumed for
the radial velocity of HCG100d.}
\end{deluxetable}

\newpage

\begin{deluxetable}{llrrc}
\tablecaption{Observational properties of the disk galaxies of the field
sample.\label{compsamp}}
\tablehead{
\colhead{Name} & 
\colhead{Type} & 
\colhead{$D$ (Mpc)} &
\colhead{$B_{0}^{T}$} & 
\colhead{$\log F_{\rm{H}\alpha}$}
}
\startdata
NGC 3623 & SAB(rs)a & 10.20 & 9.59 & -11.43 \\
NGC 4064 & SB(s)a: pec & 13.24 & 11.71 & -12.11 \\
NGC 7625 & SA(rs)a pec & 29.84 & 12.47 & -11.76 \\
NGC 2775 & Sab? & 27.01 & 10.85 & -11.63 \\
NGC 3504 & (R)SAB(s)ab & 42.21 & 11.52 & -11.40 \\
IC 356 & SA(s)ab pec & 22.13 & 10.35 & -11.83 \\
NGC 278 & SAB(rs)b & 16.27 & 10.96 & -11.17 \\
NGC 488 & SA(r)b & 38.52 & 10.83 & -12.29 \\
NGC 891 & SA(s)b? sp & 13.78 & 9.83 & -11.68 \\
NGC 1055 & SBb: sp & 17.58 & 10.79 & -12.03 \\
NGC 2683 & SA(rs)b & 8.46 & 9.84 & -11.23 \\
NGC 2841 & SA(r)b: & 16.49 & 9.58 & -11.42 \\
NGC 3351 & SB(r)b & 10.74 & 10.18 & -11.09 \\
NGC 3675 & SA(s)b & 17.69 & 10.49 & -11.21 \\
NGC 4102 & SAB(s)b & 20.78 & 11.91 & -11.72 \\
NGC 4157 & SAB(s)b? sp & 17.01 & 11.18 & -11.59 \\
NGC 157 & SAB(rs)bc & 27.99 & 10.67 & -11.11 \\
NGC 2336 & SAB(r)bc & 44.92 & 10.49 & -11.02 \\
NGC 2339 & SAB(rs)bc & 41.56 & 11.54 & -11.65 \\
NGC 2903 & SAB(rs)bc & 9.11 & 9.05 & -10.67 \\
NGC 4041 & SAB(rs)bc: & 30.05 & 11.43 & -11.25 \\
NGC 4088 & SAB(rs)bc & 19.42 & 10.60 & -11.17 \\
NGC 5248 & SAB(rs)bc & 21.48 & 10.49 & -11.15 \\
NGC 7541 & SB(rs)bc: pec & 48.03 & 11.85 & -11.60 \\
NGC 253 & SAB(s)c & 4.34 & 7.40 & -10.19 \\
NGC 628 & SA(s)c & 13.56 & 9.48 & -10.91 \\
NGC 1084 & SA(s)c & 23.44 & 10.73 & -11.14 \\
NGC 1087 & SAB(rs)c & 26.04 & 11.11 & -11.36 \\
NGC 1961 & SAB(rs)c & 72.48 & 11.15 & -11.52 \\
NGC 2276 & SAB(rs)c & 48.50 & 11.44 & -11.26 \\
NGC 2976 & SAc pec & 3.58 & 10.32 & -11.04 \\
NGC 3631 & SA(s)c & 28.64 & 10.76 & -11.13 \\
NGC 3810 & SA(rs)c & 12.48 & 10.94 & -11.15 \\
NGC 5907 & SA(s)c: sp & 18.01 & 10.08 & -11.48 \\
NGC 6643 & SA(rs)c & 33.53 & 11.07 & -11.41 \\
\enddata
\end{deluxetable}

\newpage

\addtocounter{table}{-1}

\begin{deluxetable}{llrrc}
\tablecaption{Continued.}
\tablehead{
\colhead{Name} & 
\colhead{Type} & 
\colhead{$D$ (Mpc)} &
\colhead{$B_{0}^{T}$} & 
\colhead{$\log F_{\rm{H}\alpha}$}
}
\startdata
NGC 925 & SAB(s)d & 13.56 & 10.03 & -11.10 \\
NGC 4713 & SAB(rs)d & 21.70 & 11.86 & -11.52 \\
NGC 4561 & SB(rs)dm & 21.70 & 12.44 & -11.98 \\
DDO 42 & IB(s)m & 3.5 & 11.22 & -10.87 \\
DDO 43 & Im & 7.3 & 14.72 & -13.60 \\
DDO 49 & Im: & 46.0 & 14.11 & -12.70 \\
DDO 53 & Im & 3.80 & 14.42 & -12.37 \\
DDO 218 & Im & 32 & 13.93 & -12.29 \\
NGC 3738 & Im & 5.4 & 12.06 & -12.17 \\
NGC 4214 & IAB(s)m & 5.4 & 10.16 & -11.14 \\
NGC 4449 & IBm & 5.4 & 9.86 & -10.54 \\
NGC 7800 & Im? & 39 & 12.96 & -12.26 \\
NGC 4670 & SB(s)0/a pec: & 21 & 13.01 & -11.52 \\
NGC 5253 & Im pec & 6.9 & 10.89 & -10.66 \\
Haro 22 & E1? & 28 & 12.24 & -12.67 \\
Mrk 35 & BCD/Irr & 20 & 13.11 & -11.58 \\
IV Zw 149 & (R')SAc? pec & 74 & 12.65 & -11.62 \\
NGC 3034 & I0 & 9.11 & 8.72 & -10.04 \\
NGC 3077 & I0 pec & 3.36 & 10.28 & -11.22 \\
NGC 1156 & IB(s)m & 9.66 & 11.73 & -11.64 \\
NGC 2366 & IB(s)m & 6.29 & 11.03 & -11.17 \\
NGC 2415 & Im? & 68.35 & 12.78 & -11.74 \\
Holmberg I & IAB(s)m & 6.84 & 13.01 & -12.57 \\
Holmberg II & Im & 7.60 & 11.12 & -11.46 \\
Holmberg IX & Im & 4.23 & 14.08 & -13.04 \\
\enddata
\end{deluxetable}


\newpage

\begin{thebibliography}{}

\bibitem{} Allam S., Assendorp R., Longo G., Braun M., Richter G., 1996,
\aaps, 117, 39

\bibitem{} Barnes J.E., 1989, Nature, 338, 123

\bibitem{} Burstein D., Heiles C., 1984, \apjs, 54, 33

\bibitem{} Bushouse H.A., 1987, \apj, 320, 49

\bibitem{} de Vaucouleurs G., de Vaucouleurs A., Corwin H.G., Buta R.J.,
Paturel G., Fouqu\'{e} P., 1991, Third Reference Catalog of Bright
Galaxies (Springer, New York)

\bibitem{} Ebeling H., Voges W., Boehringer H., 1994, \apj, 436, 44

\bibitem{} Giuricin G., Bertotti G., Mardirossian F., Mezzetti M., 1990,
\mnras, 247, 444

\bibitem{} G\'{o}mez-Flechoso M.A., 1997, Ph.D. Thesis, Universidad
Aut\'{o}noma de Madrid

\bibitem{} Hashimoto Y., Oemler A., Lin H., Tucker D.L., 1998, \apj,
499, 589

\bibitem{} Hickson P., 1982, \apj, 255, 382 

\bibitem{} Hickson P., Menon T.K., Palumbo G.G.C., Persic M., 1989,
\apj, 341, 679

\bibitem{} Huchra J.P., Burg R., 1992, \apj, 393, 90

\bibitem{} Huchtmeier W.K., 1997, \aap, 325, 473

\bibitem{} Hunter D.A.,  Gallagher J.S., 1986, \pasp, 98, 5

\bibitem{} Iglesias-P\'{a}ramo J., V\'{\i}lchez J.M., 1997a, \apj,
479, 190

\bibitem{} Iglesias-P\'{a}ramo J., V\'{\i}lchez J.M., 1997b, \apjl, 489,
L13

\bibitem{} Iglesias-P\'{a}ramo J., V\'{\i}lchez J.M., 1998, \aj, 115, 1791

\bibitem{} Kennicutt R.C., 1983, \apj, 272, 54

\bibitem{} Kennicutt R.C., Kent S.M., 1983, \aj, 88, 1094

\bibitem{} Kennicutt R.C., Keel W.C., van der Hulst J.M., Hummel E.,
Roettiger K.A., 1987, \aj, 93, 1011

\bibitem{} Kennicutt R.C., Tamblyn P., Congdon C.E., 1994, \apj, 435, 22

\bibitem{} Kraan-Korteweg R.C., 1986, \aaps, 66, 255

\bibitem{} Larson R.B., Tinsley B.M., 1978, \apj, 219, 46

\bibitem{} Laurikainen E., Moles M., 1989, \apjl, 345, L176

\bibitem{} Mamon G.A., 1995, ``Groups of Galaxies'', ASP Conf. Ser., 70,
83

\bibitem{} Mazzarella J.M., Bothun G.D., Boroson T.A., 1991, \aj, 101, 2034

\bibitem{} Mendes de Oliveira C., Hickson P., 1994, \apj, 427, 684

\bibitem{} Menon T.K., 1995, \mnras, 274, 845

\bibitem{} Mihos J.C., Richstone D.O., Bothun G.D., 1991, \apj, 377, 72

\bibitem{} Mihos J.C., Dubinski J., Hernquist L., 1998, \apj, 494, 183

\bibitem{} Miller B.W., Hodge P., 1994, \apj, 427, 656

\bibitem{} Moles M., del Olmo A., Perea J., Masegosa J., M\'{a}rquez I., Costa V., 1994, \aap, 285, 404

\bibitem{} Moore B., Lake G., Katz N., 1998, \apj, 495, 139

\bibitem{} Olson K.M., Kwan J., 1990, \apj, 361, 426

\bibitem{} Ostriker J.P., Lubin L.M., Hernquist L., 1995, \apjl, 444, L61

\bibitem{} Perea J., del Olmo A., Verdes-Montenegro L., Yun M.S., 1997,
\apj, 490, 166

\bibitem{} Pildis R.A., Bregman J.N., Shombert J.M., 1995, \aj, 110, 1498

\bibitem{} Ponman T.J., Bourner P.D.J., Ebeling G.H., Bohringer H.,
1996, \mnras, 283, 690

\bibitem{} Rieke G.H., Lebofsky M.J., 1985, \apj, 288, 618

\bibitem{} Saracco P., Ciliegi P., 1995, \aap, 301, 348

\bibitem{} Sulentic J.W., de Mello Raba\c{c}a D.F., 1993, \apj, 410, 520

\bibitem{} Telesco C.M., 1988, \araa, 26, 343

\bibitem{} Venugopal V.R., 1995, \mnras, 277, 455

\bibitem{} Verdes-Montenegro L., Yun M.S., Perea J., del Olmo A., Ho
P.T.P., 1998, \apj, 497, 89

\bibitem{} V\'{\i}lchez J.M., Iglesias-P\'{a}ramo J., 1998a, \apjs, 117, 1

\bibitem{} V\'{\i}lchez J.M., Iglesias-P\'{a}ramo J., 1998b, \apjl, 506,
L101

\bibitem{} Walke D.G., Mamon G.A., 1989, \aap, 225, 291

\bibitem{} Williams B.A., Rood H.J., 1987, \apjs, 63, 265

\bibitem{} Young J.S., Allen L., Kenney J.D.P., Lesser A., Rownd B.,
1996, \aj, 112, 1903

\bibitem{} Zabludoff A.I., Huchra J.P., Geller M.J., 1990, \apjs, 74, 1

\end{thebibliography}
\end{document}